\documentclass{emulateapj}
\usepackage{url}
\usepackage{epsf}

\shorttitle{Progenitor of SN~2011dh}
\newcommand{\Ni}{{$^{56}$Ni}}

\def\mean#1{{\langle}#1{\rangle}}

\shortauthors{O. G. Benvenuto, M. C. Bersten and K. Nomoto}

\begin{document} 

\title{A Binary Progenitor for the Type IIb Supernova 2011dh in M51}

\author{Omar G. Benvenuto\altaffilmark{1}\nonumber\footnote{OGB is member of the Carrera del 
Investigador Cient\'{\i}fico de la Comisi\'on de Investigaciones
Cient\'{\i}ficas de la Provincia de Buenos Aires (CIC),
Argentina.}, Melina C. Bersten\altaffilmark{2} and Ken'ichi
Nomoto\altaffilmark{2}}

\affil{\altaffilmark{1} Facultad de Ciencias Astron\'omicas y
  Geof\'{\i}sicas, Universidad Nacional de La Plata, Paseo del Bosque
  S/N, B1900FWA La Plata, Argentina.} 

\affil{\altaffilmark{2} Kavli Institute for the Physics and Mathematics of
  the Universe, Todai Institutes for Advanced Study, University of
  Tokyo, 5-1-5 Kashiwanoha, Kashiwa, Chiba 277-8583, Japan}
\email{obenvenu@fcaglp.unlp.edu.ar}
\submitted{Submitted to ApJ on June 22, 2012; accepted October 27, 2012.}

\begin{abstract} 
We perform binary stellar evolutionary calculations following
the simultaneous evolution of both stars in the system to study a
potential progenitor system for the Type IIb supernova
2011dh. Pre-explosion photometry as well as light-curve
modeling have provided constraints on the physical properties of the
progenitor system. Here we present a close binary system that
is compatible with such constraints. The system is formed by stars of
solar composition with 16~$\mathrm{M_{\odot}}$~+~10~$\mathrm{M_{\odot}}$
on a circular orbit with an initial period of 125~days. The primary star
ends its evolution as a yellow supergiant with a mass of $\approx 4
\, \mathrm{M_{\odot}}$, a final hydrogen content of $\approx 3-5 \times
10^{-3}$~$\mathrm{M_{\odot}}$ and with an effective temperature and luminosity
in agreement with the HST pre-explosion observations of
SN~2011dh. These results are nearly insensitive to the adopted
accretion efficiency factor $\beta$. At the time of 
explosion, the companion star has an effective temperature of 22 to 40
thousand Kelvin, depending on the value of $\beta$, and lies near the
zero age main sequence. Considering the uncertainties in the HST
pre-SN photometry the secondary star is only marginally detectable in
the bluest observed band. Close binary systems, as opposed to 
single stars, provide a natural frame to explain the properties 
of SN~2011dh. \end{abstract}


\section{INTRODUCTION} \label{sec:intro} 

Core-collapse supernovae (CCSNe) are the explosive end of massive
stars with  $M_{\mathrm{ZAMS}} \gtrsim 8 \mathrm{M_{\odot}}$. There is a
diversity in the spectroscopic and photometric properties of CCSNe
which are mainly related to the ability of the progenitor 
to retain its outermost layers. Type II SNe, with clear H lines in
their spectra, represents the case where a thick H envelope is kept before the
explosion.  Type Ib SNe, 
with no H lines but with clear He lines, have lost their H envelope
but not the He layers. Finally Type Ic SNe, with no H and He lines in the spectra,
represent a more extreme case where not only the the H but also the He envelopes are
likely lost before the explosion. There are also transitional objects
between these different types. One example is 
that of Type IIb SNe, which show H lines at early times but then the
spectrum is transformed into that of typical SNe~Ib \citep[see][for a
classification scheme]{1997ARA&A..35..309F}. Type IIb, Ib and Ic
objects are collectively called striped-envelope SNe \citep{1996ApJ...462..462C}.  

Progenitor models of SNe~IIb comprising a helium star surrounded by a very thin hydrogen-rich 
envelope (of $\lesssim 1 \, \mathrm{M_{\odot}}$) have been successful to explain the
observed light curves (LC) and spectral features 
\citep{1994ApJ...420..341S,1994ApJ...429..300W,1998ApJ...496..454B}.
However, it is not clear which is the mechanism responsible for the
removal of the outer envelope before the explosion. One possibility is
strong winds that occur in massive stars with 
$M_{\mathrm{ZAMS}} \gtrsim 25 \mathrm{M_{\odot}}$. Alternatively, in close binary
systems (CBS) stars are expected to exchange mass providing an
efficient mechanism to allow for the removal of outer layers. Currently, the binary channel is
favored particularly for the case of SNe~IIb
\citep{2008MNRAS.384.1109E,2011A&A_528A_131C,2011MNRAS.412.1522S}. 

Additional support for the  binary scenario in SNe~IIb   
comes from  the detection of a hot companion for the famous SN~IIb
1993J~\citep{2004Natur_427_129M}. This was initially suggested 
by pre-explosion photometry \citep{1994AJ....107..662A}. The LC
and evolutionary models of  SN~1993J were also in favor of the binary channel
\citep{1993Natur_364_507N,1993Natur_364_509P,1994ApJ...429..300W}.    
Some evidence for a companion was also reported for another SN~IIb 2001ig
\citep{2006MNRAS.369L..32R}. 

The Type IIb SN~2011dh was recently discovered in the nearby galaxy
M51 attracting the attention of many observers because of its proximity
and brightness. It was discovered almost immediately after  explosion
\citep{2011ApJ_742L_18A}. It showed early radio and X-ray 
emission \citep{2012ApJ...752...78S}. Using pre-explosion images
obtained from the HST archive 
\citet{2011ApJ_739L_37M} and \citet{2011ApJ_741L_28V} detected a
source at the location of SN~2011dh. They derived similar values of luminosity and
effective temperature for pre-SN source.
The object was consistent with a yellow supergiant (YSG) star with a 
radius $R \approx 270 \, \mathrm{R_{\odot}}$ and without any clear evidence of a
companion star contributing to the observed spectral energy distribution (SED).

At present there is a controversy in the literature as to whether the
YSG is the actual progenitor of SN~2011dh. Some authors have suggested that the
exploding star should be more compact
\citep{2011ApJ_742L_18A,2012ApJ...752...78S,2011ApJ_741L_28V} based 
on (1) a simple comparison between the early light curve (LC) of
SN~2011dh and SN~1993J; (2) a discrepancy  between the temperature derived from an early-time
spectrum and that predicted by an analytic expressions for an extended progenitor; and
(3) the large shock velocity derived from radio observations. 

Recently, we have performed a detailed hydrodynamical
modeling of SN~2011dh using stellar evolutionary progenitors
\citep{2012ApJ...757...31B}.  
These models indicate that observations are 
compatible with a helium star progenitor of a mass near
$4$~$\mathrm{M_{\odot}}$
surrounded by a thin hydrogen-rich envelope ($\approx$ $0.1$
$\mathrm{M_{\odot}}$) with a radius of  
$\approx 200 \,\mathrm{R_{\odot}}$ (similar to that of the detected YSG star)
that underwent an explosion with an energy of $8\times10^{50}$~erg
that synthesized $0.063$~$\mathrm{M_{\odot}}$ of \Ni. 
Such large radius values are needed to reproduce the early light curve
of SN~2011dh without contradicting the temperatures derived from the spectra. In
addition, our hydrodynamical modeling rules out progenitors with He
core masses larger than 8~$\mathrm{M_{\odot}}$, which corresponds to 
$M_{\mathrm{ZAMS}} \gtrsim 25 \mathrm{M_{\odot}}$.  

It is very difficult for a single star to reach these pre-SN
conditions. The existence of a strong wind capable of removing most of
the envelope requires a massive star of $\approx$ 25 $\mathrm{M_{\odot}}$ or
more \citep{2003ApJ...591..288H,2009A&A...502..611G}, which is in
contradiction  with the LC models.   
Moreover, in order to retain a thin hydrogen-rich layer, the mass loss 
rate would have to be on a very narrow interval. These facts strongly
suggests that the progenitor of SN~2011dh should be a component of a binary system.

However, a recent work by
\citet{2012A&A_538L_8G} proposed that single YSG stars such as the one
detected at the location of SN~2011dh are plausible SN
progenitors. This is based on stellar evolution calculations of stars
with main sequence masses of 12--15 $\mathrm{M_{\odot}}$ under the assumption of
an increased mass-loss rate several times above the standard
values. However, no physical explanation is given for such an
increase. Also note that a recent paper by \citet{2011A&A...526A.156M} found a 
good agreement between modern determinations of
mass-loss for RSGs and the standard mass-loss prescription
\citep{1988A&AS_72_259D}.  Although, other mass-loss formulation
as proposed by \citet{2005A&A...438..273V} point towards higher mass-loss 
rates but  this prescription seems
to be applicable only to dusty stars and gives a  overestimates of the
mass-loss rates for Galactic RSGs.
 
The aim of this work is to show the plausibility that the progenitor
of SN~2011dh was part of a close binary system (CBS) with
properties compatible with the pre-SN observations and the
results of LC modeling. The observational properties of the
remaining companion star are discussed in anticipation of future
detections. Although we do not perform a 
complete exploration of the parameter space (stellar masses, initial
orbital period, and mass-transfer efficiency $\beta$), we show that
our results are robust if we consider moderate changes of the initial
conditions.

The remainder of this paper is organized as follows. In
Section~\ref{sec:code} we  
present a brief description of our binary stellar evolution code paying special 
attention to the characteristics that enabled us to compute pre-SN
models. In Section~\ref{sec:results} we present the main results of
this paper regarding the adopted binary configuration
(\S~\ref{subsec:configuration}), evolutionary calculations 
(\S~\ref{subsec:evolu}) as well as the spectra of the components at
the moment of the explosion (\S~\ref{subsec:sed}). In
Section~\ref{sec:disc} we present a 
discussion of  our results and finally, in
Section~\ref{sec:conclu} we provide some concluding remarks. 

\section{THE STELLAR CODE} \label{sec:code}

In order to compute the evolution of the CBSs quoted above we shall
employ a code 
similar to that described in~\citet{2003MNRAS_342_50B}, adapted for
the case of 
massive stars. Briefly, this is a Henyey code that when the star
reaches Roche Lobe Over Flow (hereafter RLOF) conditions it solves
implicitly not only the whole structure of the donor star 
but also the Mass Transfer Rate (hereafter MTR)~$\dot{M}_{1}$ in a
simultaneous, fully implicit 
way. Such a procedure has been found to largely improve numerical
stability as 
compared to algorithms that compute the MTR
explicitly~\citep{2006A&A_445_647B},  
enabling us to reach pre-SN condition meanwhile the donor star is
losing mass at an appreciable 
rate with a detailed, continuous and convergent sequence of stellar
models. In detached 
conditions, the code works as a standard Henyey scheme.

In order to adapt the code to the case of massive stars we have
incorporated several 
nuclear reactions from the compilation of~\citet{1988ADNDT_40_283C}
and rewritten 
the difference equations to largely improve numerical stability. Also, we
incorporated semiconvection following the diffusive approach presented
in~\citet{1983A&A_126_207L} and first applied to massive stars
by~\citet{1985A&A_145_179L}. This formulation of diffusive
semiconvection depends 
on an efficiency parameter $\alpha_{SC}$ for which we assumed
$\alpha_{SC}= 1$ as in~\citet{2010ApJ_725_940Y}. 
In this paper we shall ignore the effects due to overshooting and
rotation. 

In semidetached conditions we compute~$\dot{M}_{1}$ following the
prescription given by~\citet{1988A&A_202_93R} 

\begin{equation}
\dot{M_{1}}= - \dot{M_{0}} \exp{\bigg(\frac{R_{1}-R_{L}}{H_{p}}\bigg)}
\label{eq:mdot}
\end{equation}

where $R_{1}$ is the radius of the 
donor star; $R_{L}$, the radius of a sphere with a volume equal to the
corresponding Roche lobe is computed following~\citet{1983ApJ_268_368E};
$\dot{M_{0}}>0$ is a smooth function of $M_{1}$ and $M_{2}$  whereas
$H_{p}=-dr/d\ln{p}$ (where $r$ and $p$ are the radial coordinate and
pressure respectively) is the photospheric pressure scale height. For
further 
details see~\citet{1988A&A_202_93R}. In detached conditions we consider
stellar wind following~\citet{1988A&AS_72_259D}. As we shall discuss 
in~\S~\ref{sec:results}, we consider different accretion
  efficiencies, $\beta$, of the material transferred by the donor to the
  secondary, i.e. $\dot{M}_{2}= -\beta \dot{M}_{1}$, and evolve the
orbit as described in~\citet{2003MNRAS_342_50B}.   

After computing the evolution of the donor star, we evolve the companion
star taking into account the accretion rate it receives from the donor
as described in~\citet{1977PASJ_29_249N}. We do not consider mass loss
from the accreting star. As we ignore any effect of the secondary star
on the donor one other than imposing a limit on its volume, this is an
usual assumption in this type of problems. 

It is well known that in CBSs like those we are interested in here,
the material accreted by the secondary star may have a mean molecular weight higher
than the corresponding to its outer layers. This may lead to an unstable
situation that induces the so~-~called thermohaline mixing~\citep{1980A&A_91_175K}. In the
present version of our code we have not included thermohaline mixing
yet. In any case, it is worth to remark that
\citet{2009MNRAS.396.1699S} have shown that ignoring thermohaline 
mixing has a minor effect on the evolutionary track of secondary stars
of CBSs similar to those studied here. Therefore, we expect that the main
conclusions of this work not to be affected by neglecting this
phenomenon. 
 
\section{THE BINARY MODELS FOR SN 2011dh}
\label{sec:results}

\subsection{The initial configuration}
\label{subsec:configuration}

The  broad-band HST pre-explosion photometry of the SN~2011dh
\citep{2011ApJ_739L_37M,2011ApJ_741L_28V} and the light
curve modeling~\citep{2012ApJ...757...31B} impose strong constraints about our
election of the binary parameters, mass of the primary, of the secondary
and the initial period. Here we briefly discuss the
motivations of our election of these parameters before we present our results. 

If the object detected in the pre-explosion image is indeed
the progenitor of SN~2011dh and it belongs to a binary system, the
primary (donor)  star needs to have a luminosity ($L$) compatible with
the value derived by \citet{2011ApJ_739L_37M} and
\citet{2011ApJ_741L_28V} for the progenitor candidate, i.e. 
$\log{L/\mathrm{L_\odot}} = 4.92 \pm 
0.2$. This luminosity is an indication of the He core mass at the
moment of the explosion which in turn is related with the mass of the star
on the main sequence ($M_{\mathrm{ZAMS}}$).  \citet{2011ApJ_739L_37M}
derived  $M_{\mathrm{ZAMS}}= 13 \pm 3\, \mathrm{M_{\odot}}$ comparing the
luminosity with the end points of the evolutionary tracks of single
stars. Note that they did not use the color information of
the progenitor to derive this mass because uncertainties are
expected in the color due to unknown mass-loss history.
 Meanwhile \citet{2011ApJ_741L_28V}
derived  $M_{\mathrm{ZAMS}}= 17-19 \,\mathrm{M_{\odot}}$ using both $L$ and
the effective temperature ($T_{\mathrm{eff}}$) derived from colors and
choosing  
the track that best matched these values, although this point does not
correspond with the final position of the single star at the end of the
evolution. In addition, ~\citet{2012ApJ...757...31B}
derived a helium core of $\approx 4 \mathrm{M_{\odot}}$ from the light curve
modeling of SN~2011dh and firmly ruled out progenitors with
$M_{\mathrm{ZAMS}} > $ 25 $\mathrm{M_{\odot}}$. Here we adopt an
initial mass of $16 \,\mathrm{M_{\odot}}$ for the donor star that is well
within the ranges provided in previous studies.

After assuming a mass value for the donor star, we still have to
consider the mass of the secondary star as well the orbital period of
the binary. Our choice should be guided by the fact that pre-explosion
observations indicate that the observed portion of the spectrum is
compatible with a single source. Obviously, the secondary has to be less
massive than the donor star but we have to distinguish between two
cases: i) the mass ratio is close to one and ii) the mass ratio
appreciably differs from one.

Let us consider the case i): If the masses of the stars differ in
(say) few tents of solar mass, the secondary star would be able to
exhaust core hydrogen before explosion. This 
object would receive material coming from the donor star when it is on
the Hertzsprung gap where the shell nuclear burning around the core
takes place. Calculations available in the literature  as well as  
 our own test (see \S~\ref{sec:disc}) indicate that,
after accretion, such object appears as an overluminous B supergiant
with a $T_{\mathrm{eff}}$ in between that of the observed YSG and that
of the ZAMS (see, e.g., Fig. 5 of Claeys, et al. 2011 and our
Fig.~\ref{fig:explora_qmasa}). 
In this case some evidence of the secondary should have been 
detected in the HST pre-explosion photometry as in the case of 
SN~1993J. 

Case ii) results more natural. If the secondary star has a mass
appreciably lower than that of the donor, it will be still undergoing
core hydrogen burning at the moment of the explosion of the primary
star. Therefore, at the moment of the explosion and in the HRD, 
we expect the secondary to be close to the ZAMS. The object will
remain hot and will emit most of the flux in the UV. As it is well
known the more massive the object the greater its luminosity on the
ZAMS and therefore the greater its effect on the pre-explosion
photometry. Thus, the mass value chosen for the secondary star should be low
enough to have remained almost undetected. Note, however, that if the
mass of the secondary is much lower than 
10 $\mathrm{M_{\odot}}$, the Kelvin-Helmholtz timescale increases appreciably and
it is very likely 
that the system reaches common envelope conditions. 
Therefore, we adopt a mass of 10 $\mathrm{M_{\odot}}$ for the secondary. 

Even with a mass of 10 $\mathrm{M_{\odot}}$ if the
secondary object  
were able to undergo a conservative mass transfer and the primary star
ends its evolution with 
a mass of $\approx 4 \,\mathrm{M_{\odot}}$ as required by LC modeling, the
secondary would end its evolution with $\approx 22 \, \mathrm{M_{\odot}}$ on the
ZAMS. This would produce a very bright object that may not be
compatible with the pre-SN observations. One  possibility  that we
will explore in the next section is that the secondary captures a
fraction $\beta$ of the material 
transferred, therefore its final mass will be of $\approx 10 + 12\;
\beta\; \mathrm{M_{\odot}}$. Note that the accretion efficiency,
$\beta$,  is
one  of the most uncertain parameters in binary stellar
evolution. Another parameter usually employed in the  
treatment of orbital evolution
in binaries is the specific angular momentum $\alpha$ of the material
lost by the system in units of the
angular momentum of the primary star. We shall assume  $\alpha= 1$
throughout this paper.

Regarding the initial period, this has been evaluated in order for the
pre-SN donor star to fall inside the error box given by the $L$ and
$T_{\mathrm{eff}}$ estimated for the progenitor candidate of SN~2011dh.
 Suppose that the donor star is losing mass by RLOF at the moment of
 the explosion, i.e. that the the size of the Roche Lobe is
 approximately equal to the radius of the donor star. This radius can
 be determined using  $\mathrm{\log\, L/L_\odot}= 4.92 \pm 0.2$ and
 $\mathrm{log}\, T_{\mathrm{eff}}= 3.78 \pm 0.02 $
 \citep{2011ApJ_739L_37M,2011ApJ_741L_28V}  
leading to values of $\approx $ 270 $\mathrm{R_{\odot}}$. If the initial masses are
16 + 10 $\mathrm{M_{\odot}}$ and the final masses are  4 + 12
$\mathrm{M_{\odot}}$, the final orbital semiaxis is $\approx 900 \, \mathrm{R_{\odot}}$ and
the final period at the moment of the explosion can be estimated
 to be $\approx$ 800 days. Finally, the initial period can be
 calculated using Eqs.~6-8 of \citet{2002ApJ_565_1107P} which
 relates the initial and final orbital semiaxes 
as a function of the initial and final masses of the system. This leads
to  initial periods of $\approx$ 120 days. Note that if the luminosity
 is due to the internal 
structure of the star, then at pre-SN conditions a given orbital
period will correspond to a given effective temperature: the larger
the period the lower the effective temperature.  In the next section we
analyze in detail a system with an initial period of 125 days
consistent with this first approximation. Other values are
 discussed in section~\ref{sec:conclu}.

\subsection{Evolutionary Results}
\label{subsec:evolu}

\begin{figure} \begin{center}
\includegraphics[scale=.350,angle=0]{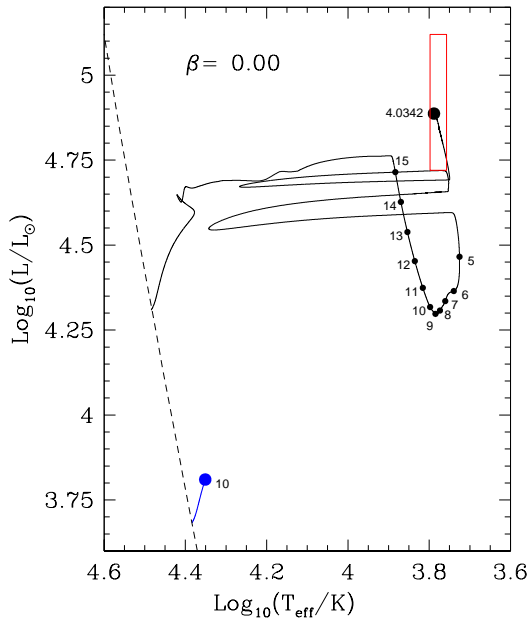} \caption{The evolution of 
the components of a close binary system of solar composition stars of
16~$\mathrm{M_{\odot}}$ and 10~$\mathrm{M_{\odot}}$ on an orbit with an initial period of 125~days
assuming fully non~-~conservative mass transfer ($\beta= 0.00$). Solid
black line represents the evolutionary 
track of the donor star. Dots along it indicate the mass of the star during the
RLOFs. The corresponding labels are in solar mass units. The star ends its
evolution with a mass of 4.034~$\mathrm{M_{\odot}}$ with an effective temperature and
luminosity compatible with the data observed for the object at the position of
the supernova  SN2011dh. Meanwhile the donor star evolves from the ZAMS
to pre-SN conditions, the companion star suffers a much slower evolution. Its
evolutionary track is depicted with a solid blue line and ends with a dot
representing the conditions attained at the moment of the explosion of the
primary. For comparison, the ZAMS corresponding to objects of the same
composition is shown in dashed
line. \label{fig:evolu_beta_00}} \end{center} \end{figure}

\begin{figure} \begin{center}
\includegraphics[scale=.350,angle=0]{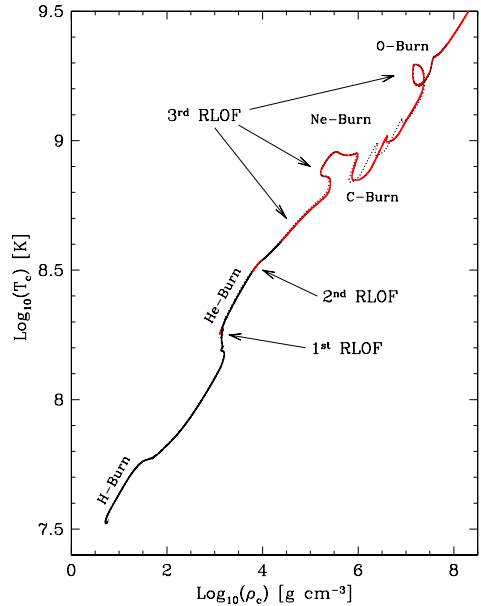} \caption{The evolution of 
the central temperature as a function of the central density. Solid line
corresponds to the evolution of the donor star for the case of $\beta=0$. Black
color indicates detached condition, while in red we depict the evolution during RLOFs. 
Notice that the first RLOF, in which three fourths of its initial 
mass are transferred, corresponds almost to a point. 
Dotted line depicts the evolution of an isolated star of the same initial mass ($16 \, \mathrm{M_{\odot}}$).
For other values of $\beta(>0)$ the
tracks are almost indistinguishable from that corresponding to $\beta=0$ and are not
included.  \label{fig:evolu_central}} \end{center} \end{figure}

\begin{figure} \begin{center}
\includegraphics[scale=.350,angle=0]{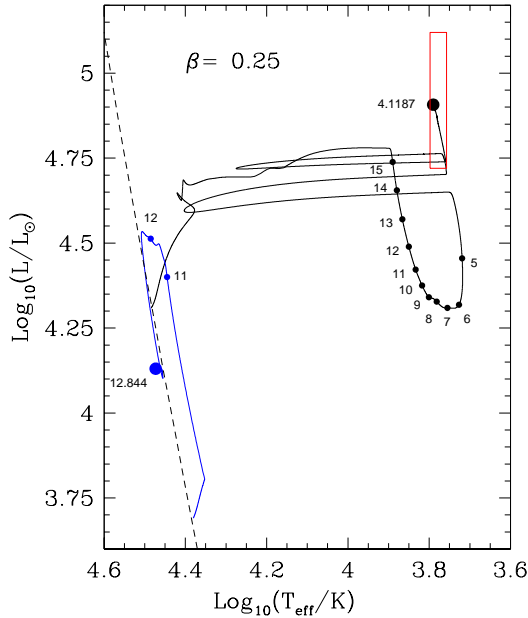} \caption{Same as
Fig.~\ref{fig:evolu_beta_00} but for the case of $\beta= 0.25$. Lines and dots
have the same meaning as there. \label{fig:evolu_beta_025}} \end{center} \end{figure}

\begin{figure} \begin{center}
\includegraphics[scale=.350,angle=0]{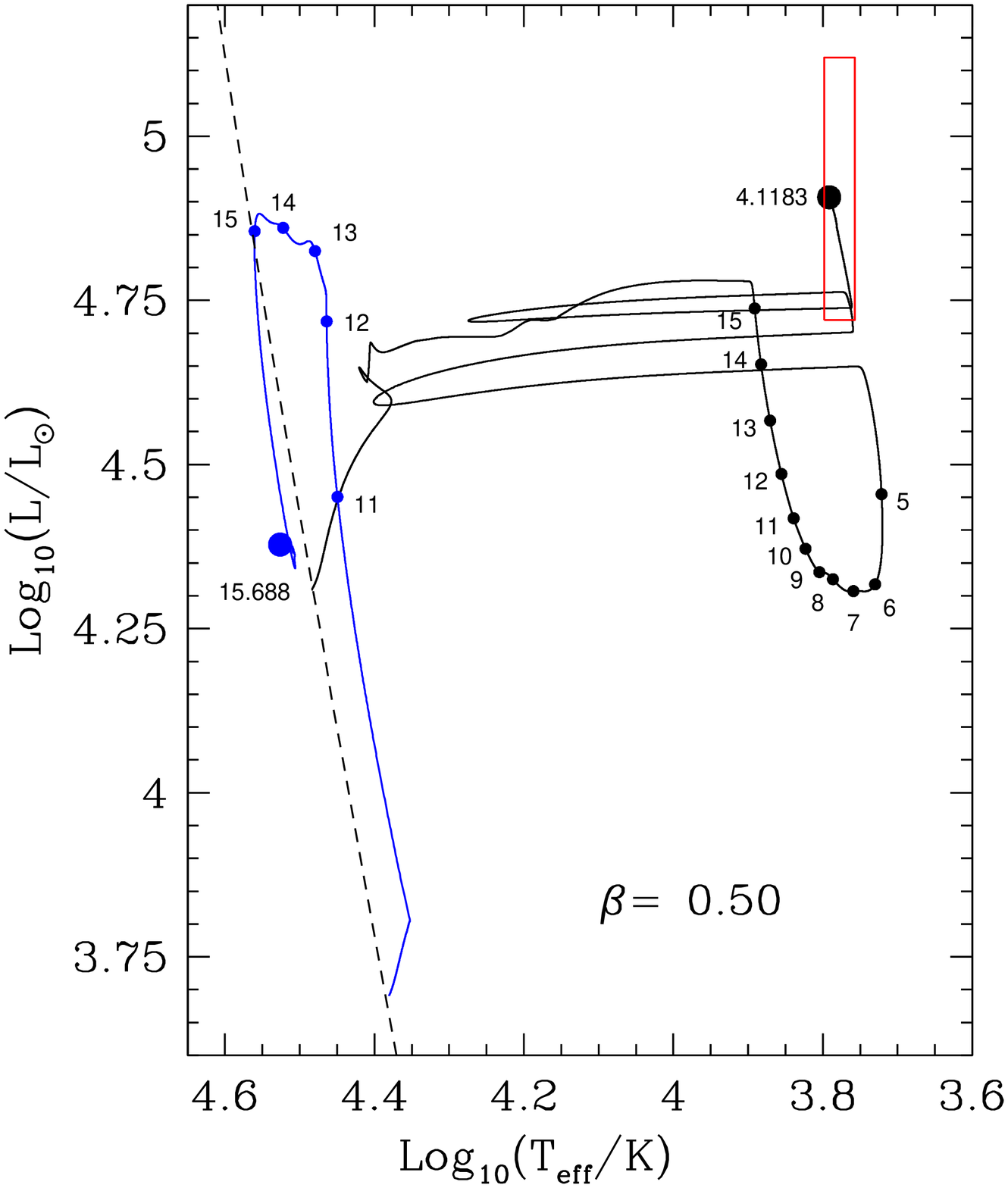} \caption{Same as
Fig.~\ref{fig:evolu_beta_00} but for the case of $\beta= 0.50$. Lines and dots
have the same meaning as there. \label{fig:evolu_beta_050}} \end{center} \end{figure}

\begin{figure} \begin{center}
\includegraphics[scale=.350,angle=0]{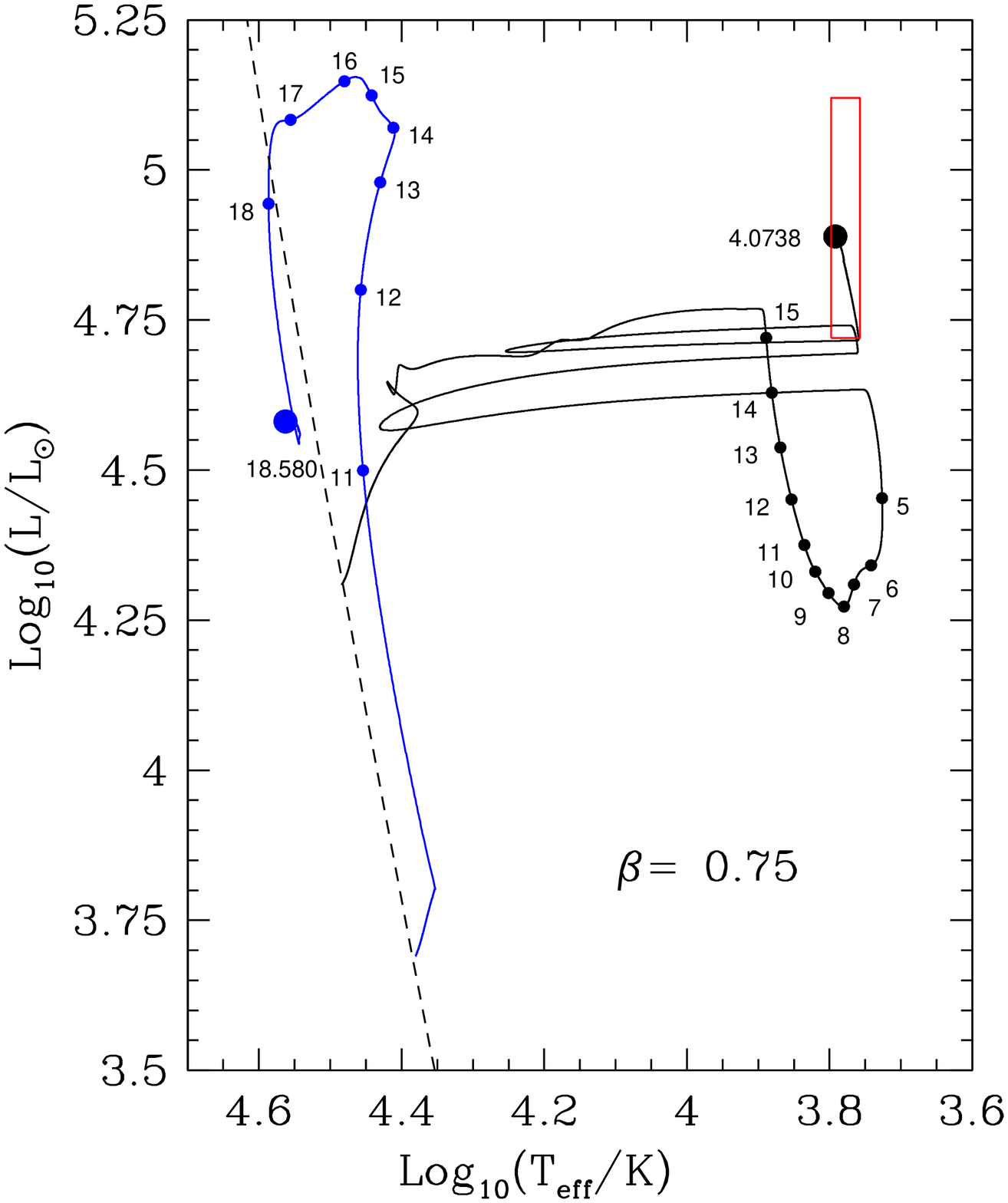} \caption{Same as
Fig.~\ref{fig:evolu_beta_00} but for the case of $\beta= 0.75$. Lines and dots
have the same meaning as there. \label{fig:evolu_beta_075}} \end{center} \end{figure}

\begin{figure} \begin{center}
\includegraphics[scale=.350,angle=0]{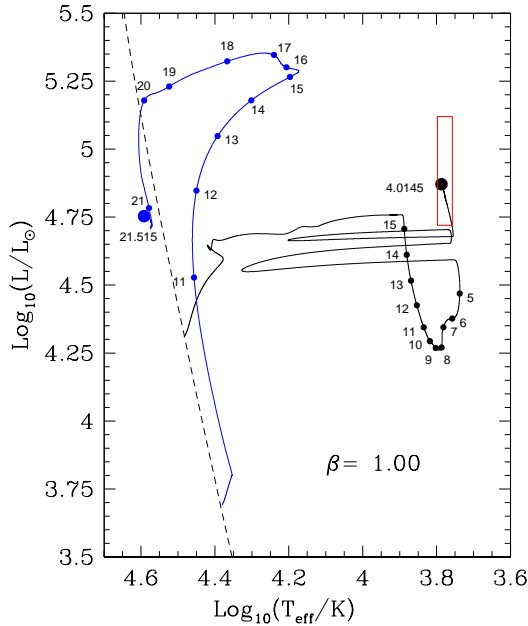} \caption{Same as
Fig.~\ref{fig:evolu_beta_00} but for the case of conservative mass
transfer ($\beta= 1.00$). Lines and dots
have the same meaning as there. \label{fig:evolu_beta_100}} \end{center} \end{figure}

\begin{figure} \begin{center}
\includegraphics[scale=.350,angle=0] {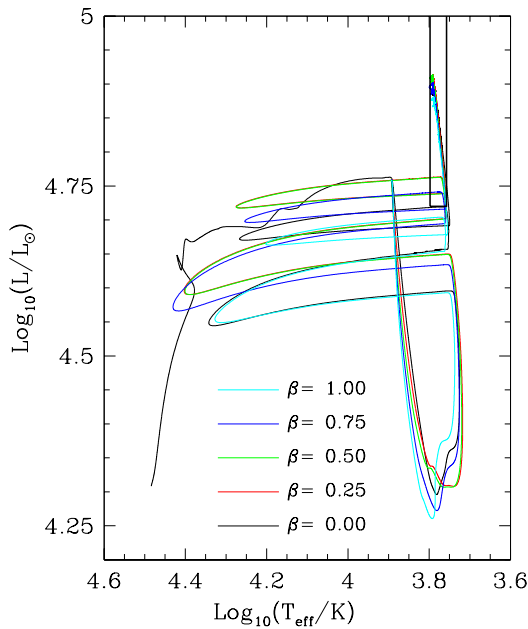} \caption{In order
to ease its comparison, we show the evolutionary tracks of the donor stars for
all the values of $\beta$ considered in this paper. Notice that all tracks are
very similar each other. \label{fig:evolu_todo_donor}} \end{center}
\end{figure}

\begin{figure} \begin{center}
\includegraphics[scale=.350,angle=0] {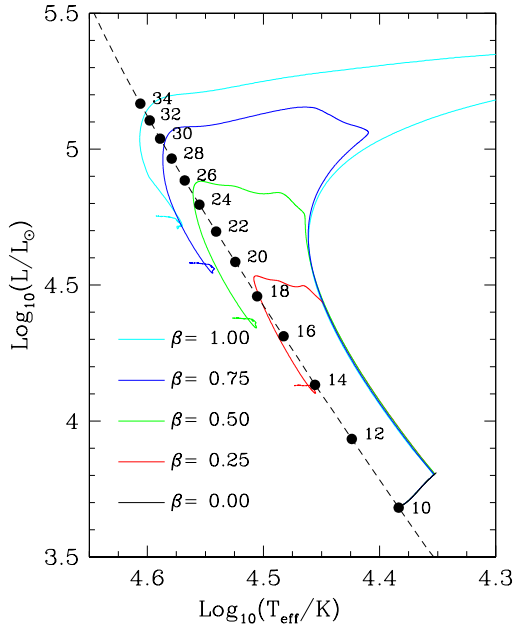} \caption{Same as
Fig.~\ref{fig:evolu_todo_donor} but for the case of the secondary
star. In sharp contrast with  
the case of the donor star, the evolution of the secondary star is
strongly dependent on the value of $\beta$. We show the ZAMS with
dashed line on which  
we have indicated some values of mass (in solar units). For the cases
of $\beta>0$  
the final position in the HRD for the accreting star is somewhat
hotter and overluminous  
than objects of the same mass on the ZAMS.  \label{fig:evolu_todo_acretor}}
\end{center} \end{figure}

\begin{figure} \begin{center}
\includegraphics[scale=.350,angle=0] {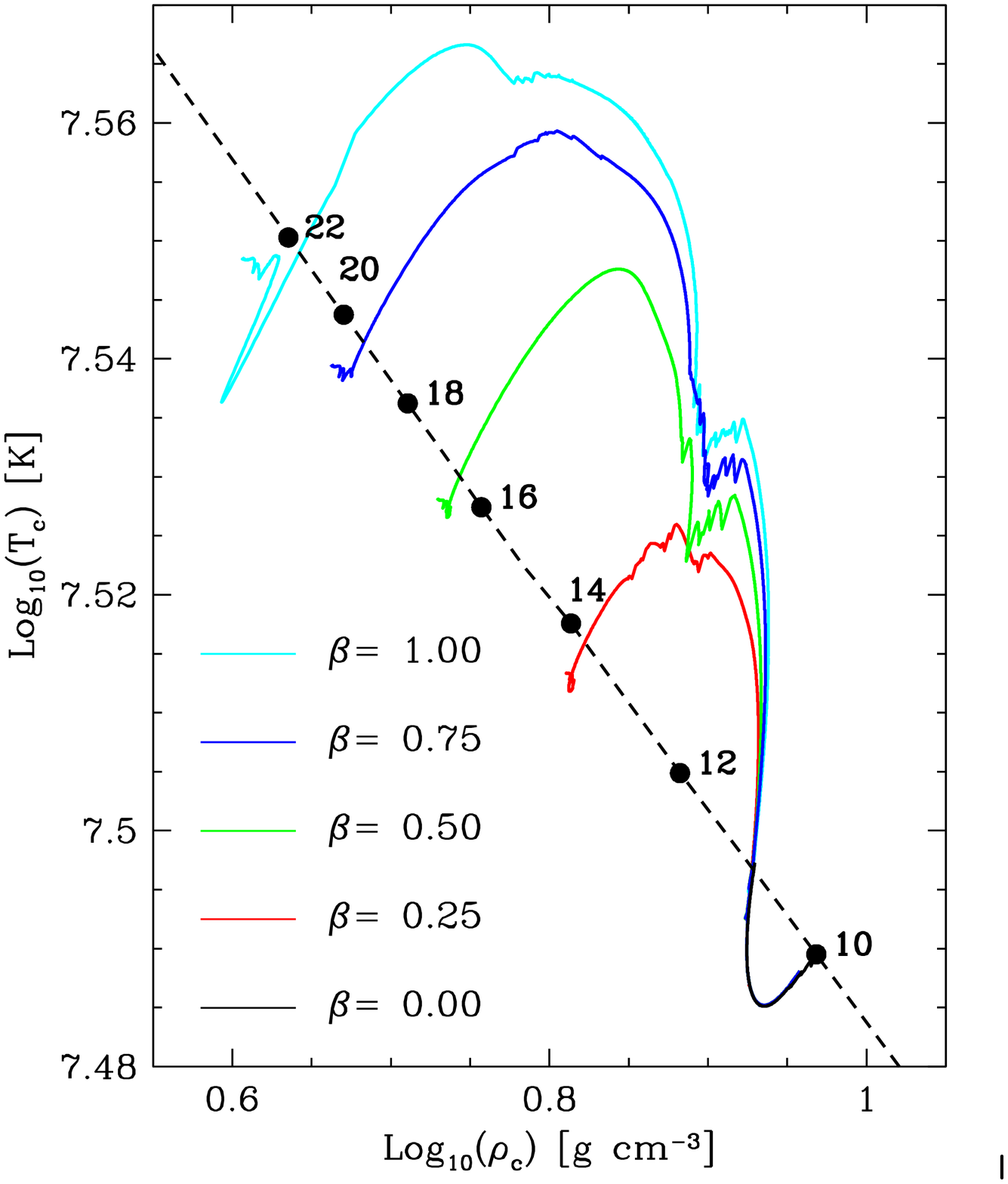} \caption{The evolution of 
the central temperature as a function of the central density for the secondary,
accreting star. This is the equivalent to Fig.~\ref{fig:evolu_central} which 
corresponds to the central evolution of the donor star but, in any case
we should remark that the scale is completely different because the
secondary star 
is still burning hydrogen during its evolution up to the explosion of
the donor star. \label{fig:centro_secund}} 
\end{center} \end{figure}

\begin{figure} \begin{center} \includegraphics[scale=.350,angle=0]
{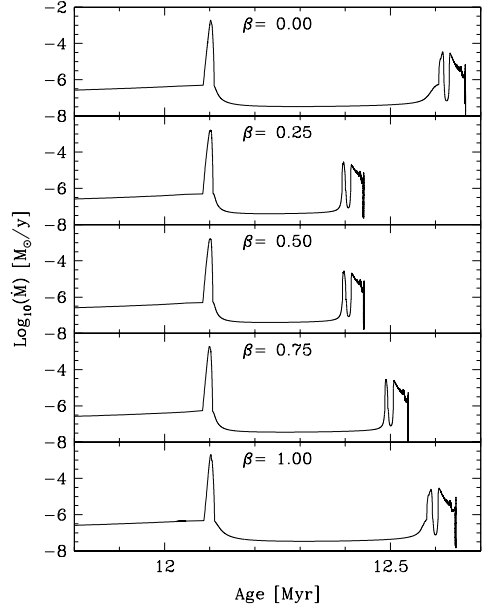} \caption{The mass transfer rate of mass from the donor star
as a function of  time. In each panel we show the cases for the five
values of $\beta$ considered in this work. In all cases there occur three
RLOFs that have a very similar profile. The only significant change is in
the time spent by the star from the end of the first RLOF to the onset of
the second one. These differences are related to changes in the orbital
semi-axis.  \label{fig:evolu_mdots}} \end{center} \end{figure}

\begin{figure} \begin{center} \includegraphics[scale=.350,angle=0]
{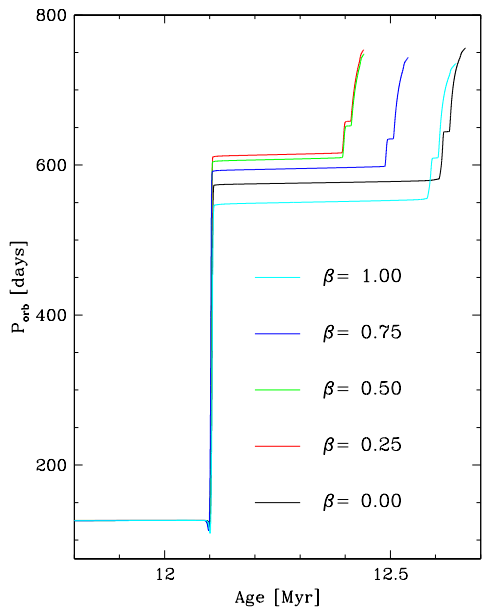} \caption{The temporal evolution of the orbital period for
the five values of $\beta$ considered in this work.   
\label{fig:evolu_orbita}} \end{center} \end{figure}

\begin{figure} \begin{center} \includegraphics[scale=.35,angle=0]
{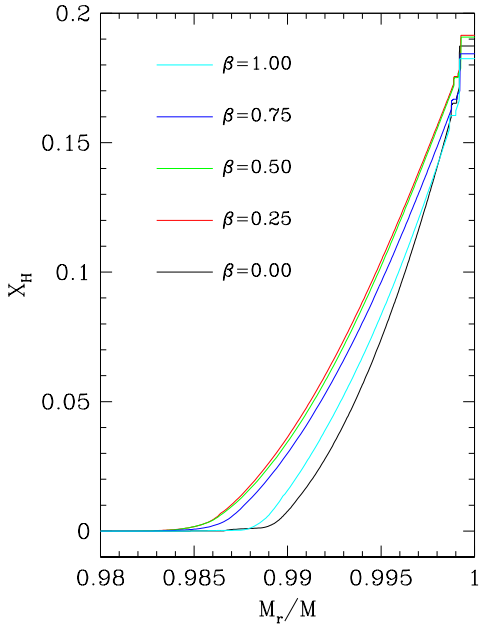} \caption{The final outermost hydrogen 
profile of pre-SN models prior to the explosion for all values of
$\beta$. Notice that these are very similar each other.
\label{fig:todo_perfilH}} \end{center} \end{figure}

In the previous section we discussed our election of the binary
parameters adopted to study a possible progenitor for SN~2011dh. Here we
present our results for a CBS of solar-composition stars with masses of
16~$\mathrm{M_{\odot}}$~+~10~$\mathrm{M_{\odot}}$ on a  circular orbit with an
initial period of 125~days. As stated
in~\S~\ref{subsec:configuration}, we analyzed
different values  of the mass-transfer efficiency,
$\beta= 0.00, 0.25, 0.50, 0.75$, and $1.00$. We computed the
evolution of both stars starting with ZAMS models up to core oxygen
exhaustion. The main results of these calculations  are presented in
Figures~\ref{fig:evolu_beta_00} to
\ref{fig:todo_perfilH}. Notice that the evolutionary tracks corresponding specifically
to the cases $\beta= 0.25$ and  0.75 are provided only in the electronic edition.

As an extreme case, we show in Fig.~\ref{fig:evolu_beta_00} the
evolution of the 
donor and companion stars assuming $\beta=0.00$ (fully
non~-~conservative mass transfer).  
This CBS undergoes class~B mass transfer: the donor star fills its
Roche lobe with  
a mass of 15.54~$\mathrm{M_{\odot}}$ well after core helium ignition
(the central abundance of helium at the onset of the first RLOF is
$X_{\mathrm{He}}= 
0.32$). With respect to the stellar mass, this is the main RLOF
because the donor 
star detaches from its lobe when has only 4.54~$\mathrm{M_{\odot}}$. Remarkably,
the RLOF takes only $7.6 \times 10^{4}$~y, which implies a mean MTR
$\mean{\dot{M}}$ of $\mean{\dot{M}}= 1.44 \times 10^{-4}\;
\mathrm{M_{\odot}/y}$. However 
the maximum  mass transfer rate $\dot{M}_{\mathrm{max}}$ is of
$\dot{M}_{\mathrm{max}}= 1.90 
\times 10^{-3}\; \mathrm{M_{\odot}/y}$ (see below for further details on MTRs). At
detaching from its Roche lobe, the donor star is still undergoing core
helium 
burning and has increased its central abundance due to semiconvective
mixing 
($X_{\mathrm{He}}= 0.39$). The physical agent that sets the end of the RLOF is
the outer layers hydrogen abundance $X_{\mathrm{H}}\big|_{S}$ that fallen down
from its  
initial value of $X_{\mathrm{H}}\big|_{\mathrm{S}}= 0.70$ to $X_{\mathrm{H}}\big|_{\mathrm{S}}= 0.49$. Thus the
stellar envelope is no longer able to support its very large radius
($192\; \mathrm{R_{\odot}}$), 
and starts a fast contraction performing a blue-wards loop on the HR
diagram. 
During this loop core helium is exhausted. Soon, due to semiconvective
mixing and nuclear  shell burning the star undergoes a second and very
brief RLOF 
followed by another (smaller as compared to the previous) blue-wards loop.
Finally, the star swells and again goes through a RLOF up to its
final explosion. During this RLOF the star undergoes carbon, neon,
oxygen and silicon 
core burning (we did not computed the silicon burning stage). During
this final RLOF 
the mass transferred from the donor is a relatively small amount of
0.27~$\mathrm{M_{\odot}}$ on $3.57 \times 10^{4}$~y ($\mean{\dot{M}}= 7.95 \times
10^{-6}\; \mathrm{M_{\odot}/y}$ and $\dot{M}_{\mathrm{max}}= 2.92 \times 10^{-5}\;
\mathrm{M_{\odot}/y}$). Notice that these MTRs are far lower than the
corresponding to 
the initial RLOF.

The evolution of the central point of the donor
  star in the temperature-density plane is presented in  
Fig~\ref{fig:evolu_central}. There, it can be clearly noticed the nuclear
activity present in the central part of the star during each RLOF. It is
remarkable that during the first RLOF central density and temperature
remain almost constant. 

In this case of $\beta=0$, the secondary star does not accrete any
material and evolves as it were an isolated object. As the main
sequence lifetime of a $10 \, \mathrm{M_{\odot}}$ star is far longer than that
corresponding to the donor star, the companion star suffers from a
very small excursion in the HRD up the moment of explosion (see
Fig.~\ref{fig:evolu_beta_00}). Let us here mention that the characteristic lifetime of isolated 
stars of $10 \, \mathrm{M_{\odot}}$ and $16 \, \mathrm{M_{\odot}}$ 
is of 23.018~Myr and 12.284~Myr respectively. 

%
%

The evolution of the components of the CBS for the cases of $\beta=
0.25, 0.50, 0.75,$ and 1.00 is shown in  
Figs.~\ref{fig:evolu_beta_025}-\ref{fig:evolu_beta_100} respectively 
(Figs.~\ref{fig:evolu_beta_025} and \ref{fig:evolu_beta_075}, corresponding
to the cases of $\beta=0.25$ and 0.75 respectively, are available in the 
electronic version of this paper). An
inspection of these figures is sufficient to realize that the
evolution of the donor star is almost independent of the value of
$\beta$ while the secondary is strongly dependent on it. In
Fig.~\ref{fig:evolu_todo_donor} we show the evolutionary tracks of the
donor  star corresponding to the five values of $\beta$ considered in
this paper. They are remarkably similar. This behavior
resembles the results found in the case of low mass
CBSs~\citep{2012MNRAS.tmp.2583D}.  

In Fig.~\ref{fig:evolu_todo_acretor} we show the evolutionary tracks
of the companion, accreting star as a 
function of $\beta$. There, for comparison, we also show the ZAMS
corresponding to the initial composition of 
these stars. For the cases of $\beta>0$ the final position in the HRD
for the accreting star is somewhat 
hotter and overluminous than objects of the same mass on the
ZAMS. This is partially due to the fact that we have neglected
thermohaline mixing in our calculations. Note that in all considered
cases, the companion star does not fill its Roche lobe and no contact
configuration is found. Thus, the CBSs studied here do not undergo any
common envelope episode.    

The evolution of the central point of the accreting star is shown in
Fig.~\ref{fig:centro_secund}. Notice that the excursion of these
objects in the density~-~temperature plane is by far smaller than the
one corresponding to the donor stars. Accreting stars are not able to
exhaust central hydrogen in the time spent by the donor star to reach
pre-SN conditions 

The evolution of the MTR from the donor star is presented in
Fig.~\ref{fig:evolu_mdots}. The only difference in the MTR evolution
as a function of $\beta$ is in the time spent by the star from the end
of the first RLOF to the onset of the second one which is minimum for
$\beta=0.50$. All the cases predict a third mass-transfer episode that
lasts until the end of the evolution and whose MTR  differs
markedly from constant. Therefore any inference on the
mass of the envelope of the donor star at the time of the explosion
should take into account the appropriate mass loss in this phase. 

By analyzing the evolution of the MTR it is possible to
understand  the evolution of the companion star.
During the first RLOF, MTR reaches very high values 
($\dot{M}\gtrsim 10^{-3}\; \mathrm{M_{\odot}/y}$). If the secondary 
star is able to efficiently accrete ($\beta > 0.1$) the material coming from 
the donor star, it may swell appreciably, reaching low effective temperatures
(see Figs.~\ref{fig:evolu_beta_025}-\ref{fig:evolu_beta_100}).
For the range of initial
periods considered here ($\approx$ 125~d) this happens during the core helium
burning of the donor star. Near the end of the first RLOF, the MTR falls down and the secondary 
star evolves (in the HRD) towards the ZAMS being overluminous. Thus, it has to
evolve to lower luminosities, compatible with its internal nuclear energy release,
keeping close to the ZAMS. 

Later on, the donor star undergoes two further RLOF episodes, reaching pre-SN
conditions on the last of them. While the MTR during the first RLOF is large
enough to force the secondary star to swell, our calculations indicate that this is {\it not} the case
in the second and third RLOFs. During these RLOFs,
MTR is about two orders of magnitude lower compared to the case of the first RLOF. 
So, the secondary star remains close to the ZAMS until the explosion
of the donor star.


The evolution of the orbital period is not very sensitive to the
value of $\beta$, as can be seen in Fig.~\ref{fig:evolu_orbita} 
(see also Table~\ref{tabla:orbita}). This
is not surprising, since \citet{2002ApJ_565_1107P} showed that the
evolution of the orbital period of CBS  
depends only slightly on the accretion efficiency. Their analytic
prediction for the evolution of the period as a function of the
initial and final masses and the $\beta$ parameter fit almost
perfectly with our numerical results. 

An important result of our calculations is that despite of the fact
that the system is in semi~-~detached conditions at
the moment of explosion, the donor star retains an appreciable amount of
hydrogen in its outermost layers as it is shown in
Fig.~\ref{fig:todo_perfilH}. The total hydrogen content is again
almost independent of the value of $\beta$ and enough to account 
for the H lines observed in the spectra of SN~2011dh
\citep{2011MNRAS.414.2985D}. While 
a full exploration of the total hydrogen content of donor 
stars at pre-SN conditions for CBSs in general is beyond the scope
of the present paper, this result 
strongly indicates that the  total hydrogen content should be a
function of the initial orbital period: the 
larger the period the larger the hydrogen content. In this sense, the
progenitor of SN~2011dh may be 
considered as a transition object.

For completeness, we present in Table~\ref{tabla:donora} the main
characteristics of the pre-SN object and 
in  Table~\ref{tabla:secundaria} those properties corresponding to the
accreting, secondary star.  

In summary, we have shown that a system  with
16 $\mathrm{M_{\odot}}$ + 10 $\mathrm{M_{\odot}}$ and an initial period of 125 days,
independently of the adopted value of $\beta$, predicts that the
primary star ends its evolution within the region of the H-R diagram
compatible with the pre-SN photometry of SN~2011dh 
\citep{2011ApJ_739L_37M,2011ApJ_741L_28V}. Furthermore, at the end of
the evolution the primary star has a mass of $\approx 4 \,\mathrm{M_{\odot}}$ and
a hydrogen content of $3-4 \times 10^{-3}\; \mathrm{M_{\odot}}$, which is
consistent with the LC modeling \citep{2012ApJ...757...31B} and the SN IIb
classification of SN 2011dh.  

\begin{deluxetable}{ccc}
\tablecolumns{3} \small \tablewidth{0pt}
\tablecaption{Some Properties of the Orbit at the Moment of
  Explosion as a function of $\beta$.\label{tabla:orbita}}
\tablehead{
\colhead{$\beta$} & 
\colhead{Semiaxis} & 
\colhead{Period} \nl 
\colhead{} & 
\colhead{$ \big[ \mathrm{R_{\odot}} \big] $} & 
\colhead{$ \big[ \mathrm{d} \big] $} \nl
}  
\startdata 
0.00 & 841.8  & 755.84 \nl  
0.25 & 894.8  & 753.49 \nl
0.50 & 937.6  & 747.94 \nl
0.75 & 976.3  & 743.09 \nl 
1.00 & 1009.0 & 735.45 \nl 
\enddata 
\tablecomments{Form left to right we tabulate the value of $\beta$;
  the orbital semiaxis and the orbital period at the moment of
  explosion. For all these cases the initial period is of 125~d and the 
  orbital semiaxis is of 311.5~$\mathrm{R_{\odot}}$.} 
\end{deluxetable} 

\begin{deluxetable}{ccccccc}
\tablecaption{Some Properties of the Donor Star at the Moment of
  Explosion as a function of $\beta$.\label{tabla:donora} }
\tablecolumns{7} \small \tablewidth{0pt}
\tablehead{
\colhead{$\beta$} & 
\colhead{$M$} & 
\colhead{$\log_{10}{T_{\mathrm{eff}}}$} & 
\colhead{$\log_{10}{L}$} & 
\colhead{$R$} & 
\colhead{$ M_{\mathrm{H}} $} & 
\colhead{$ Age $} \nl
\colhead{}        & 
\colhead{$ \big[ \mathrm{M_{\odot}} \big] $} & 
\colhead{$ \big[ \mathrm{K} \big] $} & 
\colhead{$ \big[ \mathrm{L_{\odot}} \big] $} & 
\colhead{$ \big[ \mathrm{R_{\odot}} \big] $} & 
\colhead{$ \big[ 10^{-3}\; \mathrm{M_{\odot}} \big] $} & 
\colhead{$ \big[ \mathrm{Myr} \big] $} \nl
}  
\startdata 
0.00 & 4.034 & 3.788 & 4.886 & 245.54 & 3.869 & 12.66 \nl
0.25 & 4.118 & 3.790 & 4.907 & 249.20 & 4.441 & 12.44 \nl
0.50 & 4.118 & 3.791 & 4.907 & 247.73 & 4.537 & 12.44 \nl
0.75 & 4.073 & 3.791 & 4.889 & 242.18 & 4.926 & 12.53 \nl 
1.00 & 4.014 & 3.786 & 4.871 & 242.74 & 3.465 & 12.64 \nl 
\enddata 
\tablecomments{Form left to right we tabulate the value of $\beta$;
  the mass; effective temperature; luminosity; radius, total amount of
  hydrogen; and the age of the pre-SN.}
\end{deluxetable} 

\begin{deluxetable*}{cccccccc}
\tablecolumns{8} \small \tablewidth{0pt}
\tablecaption{Some Properties of the Secondary Star at the Moment of
  Explosion as a function of $\beta$.\label{tabla:secundaria}}
\tablehead{
\colhead{$\beta$} & 
\colhead{$M$} & 
\colhead{$\log_{10}{T_{\mathrm{eff}}}$} & 
\colhead{$\log_{10}{L}$} & 
\colhead{$R$} & 
\colhead{$\log_{10}{\rho_{\mathrm{c}}}$} & 
\colhead{$\log_{10}{T_{\mathrm{c}}}$} & 
\colhead{$X_{\mathrm{H}}\big|_{\mathrm{c}}$}  \nl 
\colhead{} & 
\colhead{$ \big[ \mathrm{M_{\odot}} \big] $} & 
\colhead{$ \big[ \mathrm{K} \big] $} & 
\colhead{$ \big[ \mathrm{L_{\odot}} \big] $} & 
\colhead{$ \big[ \mathrm{R_{\odot}} \big] $} & 
\colhead{$ \big[ \mathrm{g\; cm^{-3}} \big] $} & 
\colhead{$ \big[ \mathrm{K} \big] $} & 
\colhead{} \nl
}  
\startdata 
0.00 & 10.000 & 4.351 & 3.810 & 5.324 & 0.929 & 7.497 & 0.426 \nl  
0.25 & 12.844 & 4.473 & 4.130 & 4.391 & 0.810 & 7.513 & 0.543 \nl
0.50 & 15.688 & 4.526 & 4.377 & 4.570 & 0.729 & 7.522 & 0.584 \nl
0.75 & 18.580 & 4.562 & 4.581 & 4.879 & 0.662 & 7.539 & 0.597 \nl 
1.00 & 21.515 & 4.592 & 4.753 & 5.201 & 0.606 & 7.548 & 0.613 \nl 
\enddata 
\tablecomments{Form left to right we tabulate the value of $\beta$;
  the mass; effective temperature; luminosity; radius; and the central values
  of density, temperature, and hydrogen abundance at the moment of
  explosion.} 
\end{deluxetable*}

\subsection{Pre-Supernova Spectral Energy Distribution} \label{subsec:sed}
Our model calculations predict that the primary star ends its
evolution with properties ($L$ and $T_{\mathrm{eff}}$)
compatible with those inferred for the pre-explosion object located at
the SN position \citep{2011ApJ_739L_37M,2011ApJ_741L_28V}. At the same
time, the secondary star has a luminosity that, depending on the value of
$\beta$, could be comparable with that of the primary star and therefore
produce a detectable effect on the pre-SN spectral
energy distribution (SED). In all cases the secondary star is found to
be significantly hotter than the donor star. Indeed, for high enough
effective temperatures, the effect of the secondary would only be
appreciable in the bluest available photometric band.

We thus study here the effect of the secondary on the SED of the
system for different values of $\beta$, and compare this with the HST
pre-explosion photometry. To calculate the SED for each star of the
binary system as well as the composed SED we used atmospheric models for
solar composition provided by \citet{1993sssp.book.....K}. The model
spectrum of each star was obtained by linearly interpolating at the
values of $T_{\mathrm{eff}}$ and surface gravity ($g$)
given in Tables~\ref{tabla:donora} and \ref{tabla:secundaria}.  
The observed fluxes were computed by multiplying the models by
$(R/d)^2$ where $d$ is the distance to M51 assumed to be $7.1$~Mpc
\citep{2006MNRAS.372.1735T}. The sum of the SEDs was used to  
compute synthetic photometry through the HST transmission
filters
\footnote{\url{http://www.stsci.edu/hst/wfc3/insperformance/filters/}}

\begin{figure} \begin{center}   \includegraphics[scale= 0.35, angle= 0]
{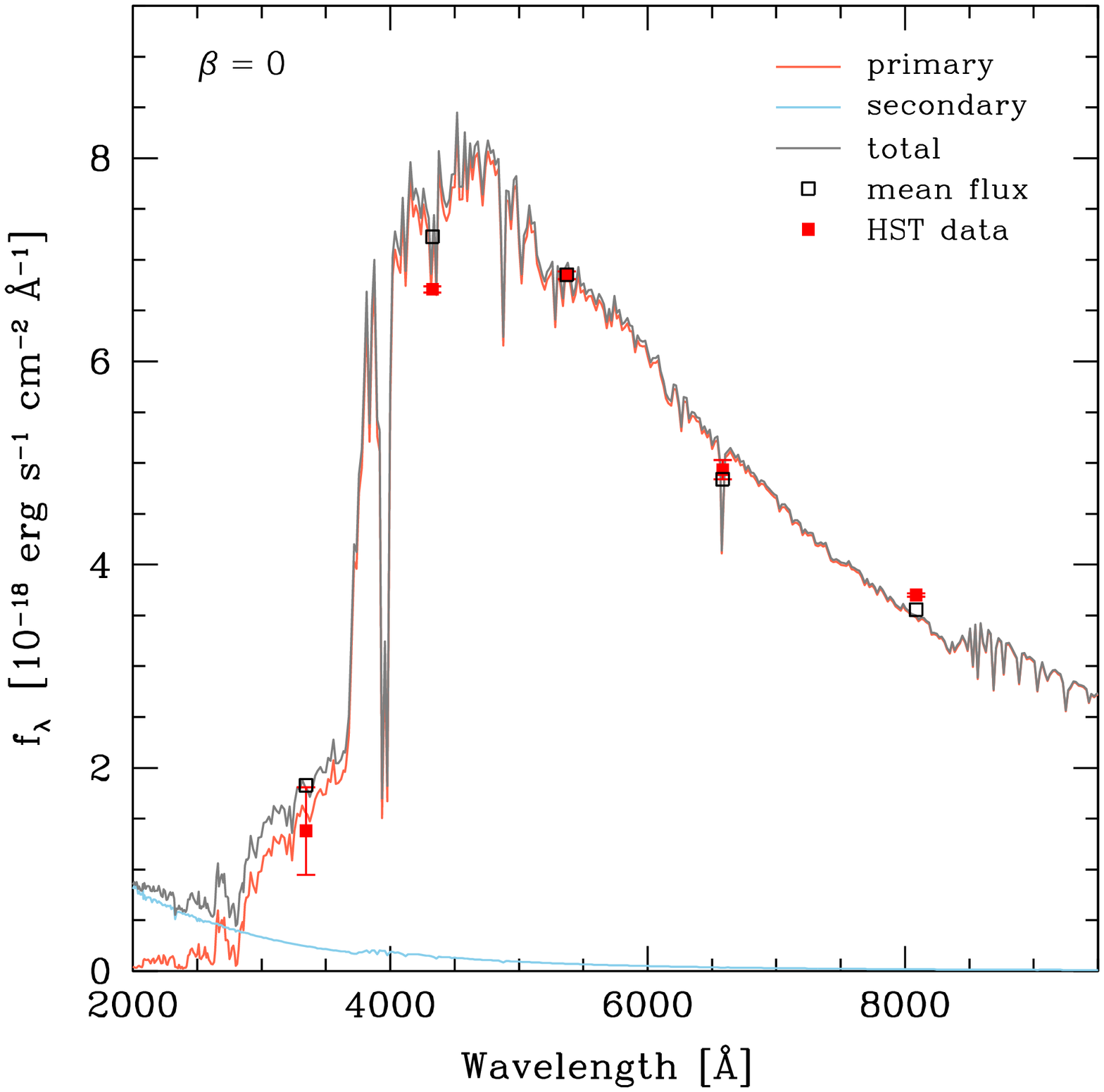}  \includegraphics[scale= 0.35, angle= 0]
{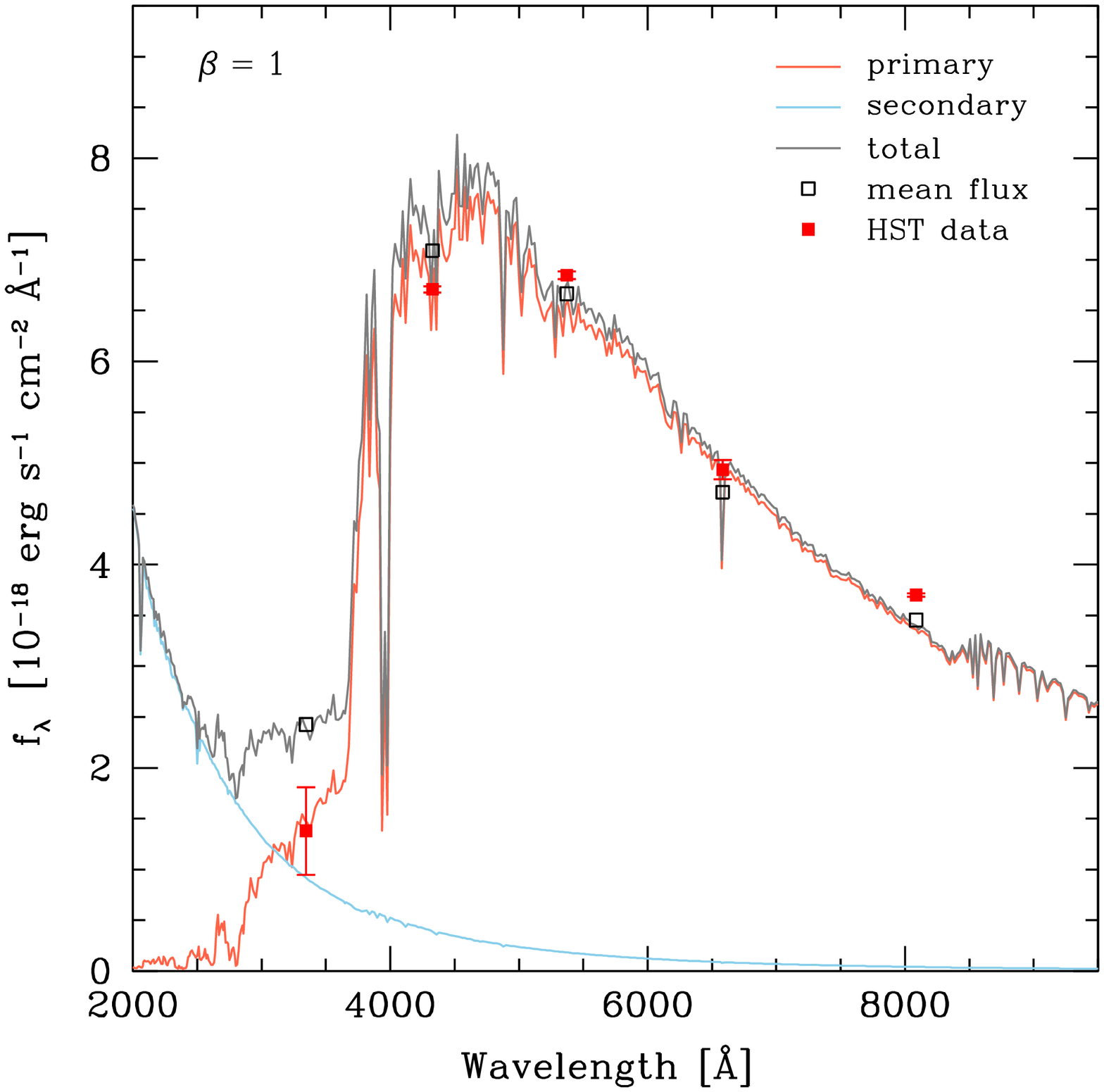}   \caption{The spectra of the donor and accreting
stars (shown with red an cyan lines respectively). Gray line represents
the addition of both spectra which, in turn, should represent the
observed one. We show the results corresponding to fully-non conservative
and conservative cases ($\beta$= 0 and 1 respectively). The mean
synthetic (black squares) and observed fluxes (red squares)  in each
bandpass are included in the figure. The secondary star is significantly
bluer than the primary at the time of explosion. Its contribution to the
total flux is non-negligible  only for the bluest observed band, F336W.
See \S~\ref{subsec:sed} for further discussion. \label{fig:espectros}}
\end{center} \end{figure}

Fig.~\ref{fig:espectros} shows the resulting SED of each star and their
sum for $\beta=$~0 and 1. 
Mean synthetic and observed fluxes are compared for all
bands. Observed fluxes were 
obtained from the tabulated magnitudes given by
\citet{2011ApJ_741L_28V} and zero points in the Vega system 
calculated using the Alpha Lyrae SED provided by
\citet{2007ASPC_364_315B}. Note that the
contribution of the secondary star is significant only for the bluest
observed band, F336W. The increase in the F336W flux due to the
presence of the secondary is of 16\%, 22\%, 32\%, 45\%, and 62\% for the cases  
of $\beta=$ 0.00, 0.25, 0.50, 0.75, and 1.00 respectively. 

For the rest of the observed HST bands the contribution of the
secondary is $<$ 6\%, i.e. for wavelengths $\lambda \geq$~4000\AA\ the
spectra is completely dominated by the light coming from the donor,
pre-SN star. In this wavelength range the
available observations  on four photometric bands    
pose no constraint on the properties of the accreting star. On the
contrary, the bluest filter, with a maximum transmittance at $\lambda
\approx$~3400 \AA\, partially detects the red tail of the spectrum of
the accreting star.

For the filters with transmittance at $\lambda \geq$~4000\AA\ the
agreement between calculations and observations  
is very good, especially considering that we have not adjusted any
parameter. For the case of the bluest filter, F336W, the contribution
of the secondary to the total flux is not negligible. If we consider the
measurement uncertainty in this band we find that the secondary is
detectable only at the 2-$\sigma$ level in the most extreme case ($\beta=1$).
The contribution is further decreased to the $0.6$-$\sigma$ level when
completely  non-conservative mass accretion ($\beta=0$) is
considered.

Remarkably, irrespective of the value of $\beta$, at the 
moment of the explosion, the secondary star is so hot
that its light has been barely detected. Thus, quite unfortunately,
the characteristics of the secondary star are very poorly constrained
by the presently available data.

\section{DISCUSSION} \label{sec:disc}
In the previous section we have shown that a properly chosen binary
configuration can explain very well the proposed YSG progenitor of
SN~2011dh. Our models also predict that the total and hydrogen masses
of the donor star at the moment of the explosion are consistent with
the results of LC modeling and the SN IIb classification.    

\begin{figure} \begin{center}
\includegraphics[scale=.350,angle=0]{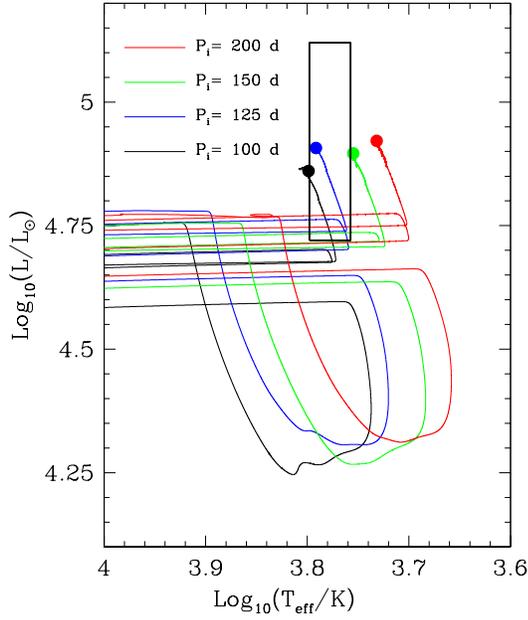}  \caption{The red
part of the evolutionary tracks for solar composition stars of binary
systems with masses of 16~$\mathrm{M_{\odot}}$ + 10~$\mathrm{M_{\odot}}$,
a fixed value of $\beta= 0.50$ and different values of the initial
orbital period. The larger the initial period the lower the effective
temperature attained by the pre-SN.   For $P_{i}= 125$~d we find the
pre-SN object inside the error box allowed by photometry previous to the
explosion without need of any fine tuning. \label{fig:explora_peri}}
\end{center} \end{figure}

A complete exploration of the parameter space of initial stellar
masses and orbital periods is not the scope of this paper.
Nevertheless, we briefly test that our
results are robust if we consider moderate changes of the initial
parameters. For example, 
Fig.~\ref{fig:explora_peri} shows the sensitivity
 of the evolutionary track of the primary star on small
variations of the initial period, $P_i$ between 100 and 200 days, for
the same configuration presented in the previous section, i.e. 16 + 10
$\mathrm{M_{\odot}}$, and $\beta=0.5$. Within such range of initial periods the
donor star ends its evolution as a YSG close to the region of the HRD
allowed by the pre-SN photometry. The evolutionary track of the secondary
is not shown in Fig.~\ref{fig:explora_peri} because it is not
sensitive to the initial period in the range under study. 
     
\begin{figure} \begin{center}
\includegraphics[scale=.350,angle=0]{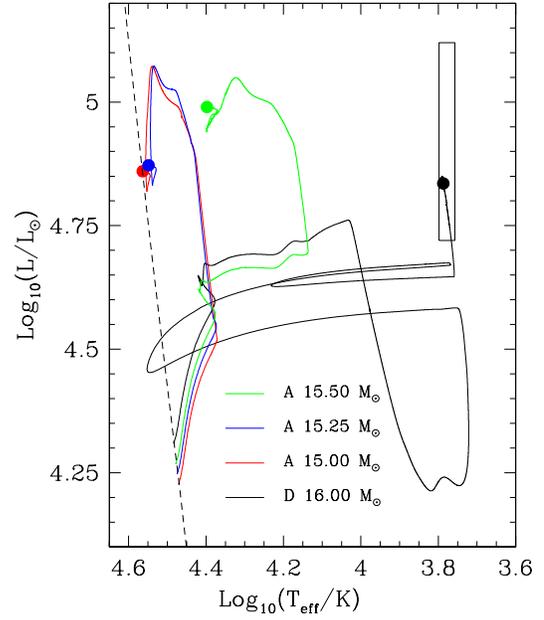}  \caption{The
evolution of binary systems with a mass of the primary of
16~$\mathrm{M_{\odot}}$, $\beta= 0.50$ and different values for the
initial mass of the companion. Labels A and D stand for accretor and
donor stars respectively. For simplicity, we computed the evolution of
the donor with the 15~$\mathrm{M_{\odot}}$ companion and assumed it to be
the same for the other companion masses. For this donor initial mass
value, secondary stars with masses up to
$\approx$~15.25~$\mathrm{M_{\odot}}$ fall close to the ZAMS at the moment
of the explosion of the donor star and most of its light will be emitted
on the blue part of the spectrum. However, if the secondary star has an
initial mass of 15.50~$\mathrm{M_{\odot}}$, at the supernova event it
will have at much lower effective temperature. This is not compatible
with the available observations of the progenitor of SN~2011dh. 
\label{fig:explora_qmasa}} \end{center} \end{figure}

The effect of changes in the mass of the secondary star on the binary
evolution is shown in Fig.~\ref{fig:explora_qmasa}. Three  
values were considered, $M_2 = 15$, $15.25$ and $15.50$ $\mathrm{M_{\odot}}$. In
all the cases, we assumed a value of $\beta=0.5$ and a mass of the
primary star of 16 $\mathrm{M_{\odot}}$ as in our previous models. The initial
period was modified to make the end-point of the donor star match the
region allowed by the pre-SN photometry. Following the criteria
described in section~\ref{subsec:configuration}, the adopted initial
period for this test was of 50 days. From the figure, it is clear that
only for the case of $M_2= 15.5 \,\mathrm{M_{\odot}}$, where the mass ratio is
closest to one, the secondary star moves appreciably away from the
ZAMS at the moment of the explosion of the primary. This is because
the secondary has exhausted its hydrogen core before the beginning
of the first RLOF. Consequently, the secondary ends with a redder color
than that of lower-mass stars and it is expected to contribute
significantly to the observed pre-SN photometry. Indeed, the
flux of the secondary in the bluest observed band, F336W, is $2.5$ times
larger than that of the primary star, which would lead to a
$8.5\,\sigma$ detection. For the rest of the HST bands the
contribution of the secondary is within 6--30$\%$. While not compatible  
with the case of SN~2011dh, this configuration may be applicable to
SN~1993J whose pre-SN observations showed evidence of a companion star
\citep{2009MNRAS.396.1699S}.

Secondary stars of slightly lower initial masses ($15$ and $15.25$
$\mathrm{M_{\odot}}$) remain near the ZAMS and end as hot stars of similar
luminosity to the primary. Stars of masses between 10 and 15
will also remain as blue objects near the ZAMS. Because of the high
effective temperature 
most of the flux from the secondary is emitted in the UV, away from
the bandpasses of the pre-explosion imaging. However, if
the luminosity is high enough, the secondary may produce a
detectable effect in the optical range. For instance, with 
masses of $15$ and $15.25$ $\mathrm{M_{\odot}}$, the flux of the secondary in the
F336W band would be comparable to that of the primary star. 
The level of detection of the secondary in these cases would be close
to $3.5\,\sigma$. The 
final luminosity of the secondary also depends on the assumed value of
$\beta$. Lower initial masses could result in similarly high
luminosities with values of $\beta$ close to unity. Unfortunately, it
is not possible to break such degeneracy based solely on the available
pre-explosion photometry. Future observations of the possible
secondary star after the SN fades from sight can shed light on this
matter.

 The initial mass of 16   
$\mathrm{M_{\odot}}$ for the primary star leads to a progenitor with the right
characteristics as derived from pre-explosion photometry, LC
modeling and spectral classification. Therefore, our choice of initial
masses and orbital period is by no means unique but it allows to prove
that a CBS is a plausible progenitor for SN~2011dh and other
SNe~IIb. Moreover, the right progenitor properties are achieved in a
self-consistent manner and independently of the detailed initial
conditions.

In the past few years, some progenitors of SNe~II have been 
associated with YSG stars, e.g. Type IIP SN~2008cn
\citep{2009ApJ...706.1174E}, Type IIL 
SN~2009kr \citep{2010ApJ...714L.280F,2010ApJ...714L.254E}, and the SN
studied here. The 
explosion of a YSG  is not compatible with the theoretical prediction
of single stellar evolution. Motivated by this apparent
discrepancy and the lack of evidence  
of a companion for SN~2011dh, \citet{2012A&A_538L_8G} studied the
effect of an increased mass-loss rate on the final properties of stars
with initial masses of 12-15 $\mathrm{M_{\odot}}$. Assuming rates  several
times higher than the standard values, they found that it was
possible to explain the explosion of single stars of relatively low
mass in the yellow area of the HRD. In particular, the proposed
scenario could explain the progenitor of SN~2011dh. However, no
physical explanation was given for such an increased mass-loss
process.

Note that the evolutionary tracks of isolated stars are strongly
dependent upon the details of mass loss. In this sense, explaining
the position of the progenitor of SN~2011dh in the HRD would
require some degree of fine tuning. The binary scenario instead provides a self-consistent
picture that naturally explains why the star remains as YSG for long 
periods of time and until the explosion. In the case of CBS evolution
we find that, irrespective of the mass transfer efficiency, the donor
star undergoes a final RLOF before igniting carbon and so it is
still transferring mass to the companion at the moment of
explosion. During the final RLOF, the donor star transfers a small
amount of material. So, the orbit and its Roche lobe enlarge very
little meanwhile the donor star tends to swell as a consequence of
nuclear shell burning. This precludes the donor star to attain lower
effective temperatures and makes it evolve to higher
luminosities in a way equivalent to the evolution of giant stars with
extended outer convective zones. In any case, the observed effective
temperature of the proposed progenitor of SN~2011dh strongly indicates
the initial orbital period, but no other parameter has to be adjusted
for the pre-SN object to fall inside the observational error box. The
YSG nature of the progenitor is thus a direct consequence of close
binary evolution.

Recently \citet{2010ApJ...711L..40C} suggested a division of 
SNe~IIb into compact (cIIb; $R \sim 10^{11}$ cm) and extended (eIIb; $R
\sim 10^{13}$ cm) subtypes essentially based on radio LC
properties. Compact objects were proposed to have smaller hydrogen
masses, roughly below $0.1$ $\mathrm{M_{\odot}}$. \citet{2011A&A_528A_131C}  
used this criterion and their own CBS evolutionary code to
analyze the range of periods (for $P_i > 1000$ days) and initial
masses needed to produce extended SNe~IIb. However, our 
calculations show that it is possible to have an extended progenitor
($R \approx 250 \, \mathrm{R_{\odot}}$) with a H mass of $< 0.1$
$\mathrm{M_{\odot}}$. Therefore, if the cIIb and eIIb subtypes
correspond to physically distinct progenitors, the division criterion
may need to be revised. In addition, \citet{2010ApJ_725_940Y}
also analyzed the space of parameters in CBSs to produce Type Ib/Ic
SNe using very different initial periods of $\lesssim$ 8 days, as
compared with the ones adopted here. They found that some of their
models predict a thin hydrogen layer of $\lesssim 0.01$ $\mathrm{M_{\odot}}$ with
a compact structure. This channel of production of SNe~IIb is
different from the one we have presented and leads to explosions far
away from the YSG regime. 
\section{CONCLUSIONS} \label{sec:conclu}

With the aim of providing a description of the progenitor of
SN~2011dh, we have studied the evolution of 
close binary systems of solar composition stars with masses of
16~$\mathrm{M_{\odot}}$~+~10~$\mathrm{M_{\odot}}$. We considered an initial period of
125~days and different efficiencies ($\beta$) of the
mass transfer process.
We followed the simultaneous evolution of the donor and
accreting stars from the zero age main sequence up to the oxygen
core exhaustion of the donor. We found that the donor star, 
independently of $\beta$, ends its evolution with  effective
temperature and luminosity consistent with the YSG object detected 
in the HST pre-SN photometry. The exploding star has a mass
$M\approx$~4\,$\mathrm{M_{\odot}}$, a radius
$R\approx$~250\,$\mathrm{R_{\odot}}$ and an outermost layer containing $3-5
\times 10^{-3}$~$\mathrm{M_{\odot}}$ of hydrogen. This is generally consistent
with the type IIb classification and the results of LC modeling 
of  SN~2011dh by \citet{2012ApJ...757...31B}. These results are a natural 
consequence of the close binary evolution and require no external
adjustment of any physical condition.

Regarding the accretion efficiency, $\beta$, we found that
(1) the evolution of the donor star 
is almost independent of $\beta$ while the secondary 
strongly depends on it, and (2) the evolution of the orbital period,
the  MTR and the total hydrogen content are almost independent of
the value of $\beta$.

Our calculations indicate that the  donor star is   lossing mass at the moment
of the explosion with rates that differ markedly from constant. Inferences
on the mass of the donor star at the time of the
explosion should take into account the appropriate mass loss in this
phase. We also found some indication that the  total hydrogen content
may be a function of the initial orbital period with larger period
producing  a larger the hydrogen content. A more detailed study of this
point is left for future work.

Note that the structure  of the donor star at the moment of the
explosion is consistent with an extended SN IIb but with very little H
mass ($< 0.1$ $\mathrm{M_{\odot}}$). 

We analyzed the effect of the secondary star on the observed HST pre-explosion
photometry. For all the values of $\beta$, at the moment of the
explosion of the donor, the secondary star is still near the
ZAMS. This is a direct consequence of our assumption that the object
has a mass appreciably lower than that of the donor. The effective
temperature of the companion is far higher than that of the 
donor with values within of 22 to 40 thousand Kelvin.  Thus, the
largest contribution to the flux of the system from the secondary is
in the bluest observed band, F336W, producing a marginal detection of       
0.6--2$\sigma$ level depending of the value of  $\beta$.
Unfortunately, the available  HST pre-SN
observations are not very suitable to constrain the properties of
the secondary.

The ultimate proof of the binary nature of SN~2011dh
must come from the possible detection of a very hot star once the SN
light fades enough. This situation would be similar to what occurred with
SN~1993J but with different properties of the companion. In any case,
we should remark that detecting the companion star of SN~2011dh would
provide valuable information on the efficiency of the mass transfer
process and evolution of massive CBSs in general. 

\acknowledgments OGB wants to thank Prof. Juan Carlos Forte for his help in
getting funds that enabled him to visit the IPMU. The funds were
provided by a PIP~712 of the Comisi\'on de Investigaciones
Cient\'{\i}ficas y T\'ecnicas (CONICET), Argentina. 
This research has been supported in part by the Grant-in-Aid for
Scientific Research of MEXT (22012003 and 23105705) and 
JSPS (23540262) and by World Premier International Research Center
Initiative, MEXT, Japan.


\clearpage


\begin{thebibliography}{}

\bibitem[Aldering et al.(1994)]{1994AJ....107..662A} Aldering, G., 
Humphreys, R.~M., \& Richmond, M.\ 1994, \aj, 107, 662  

\bibitem[Arcavi et al.(2011)]{2011ApJ_742L_18A} Arcavi, I.,Gal-Yam,
  A.,Yaron, O., et al.\ 2011, \apjl, 742, L18   

\bibitem[Benvenuto \& De Vito(2003)]{2003MNRAS_342_50B} Benvenuto,
  O.~G., \& De Vito, M.~A.\ 2003, \mnras, 342, 50   

\bibitem[Bersten et al.(2012)]{2012ApJ...757...31B} Bersten, M.~C., 
Benvenuto, O.~G., Nomoto, K., et al.\ 2012, \apj, 757, 31  

\bibitem[Blinnikov et al.(1998)]{1998ApJ...496..454B} Blinnikov, S.~I., 
Eastman, R., Bartunov, O.~S., Popolitov, V.~A.,  
\& Woosley, S.~E.\ 1998, \apj, 496, 454 

\bibitem[Bohlin(2007)]{2007ASPC_364_315B} Bohlin, R.~C.\ 2007, The Future 
of Photometric, Spectrophotometric and Polarimetric Standardization, 364, 315  

\bibitem[B{\"u}ning \& Ritter(2006)]{2006A&A_445_647B} B{\"u}ning, A., \&
Ritter, H.\ 2006, \aap, 445, 647  

\bibitem[Caughlan \& Fowler(1988)]{1988ADNDT_40_283C} Caughlan, G.~R., \&
Fowler, W.~A.\ 1988, Atomic Data and Nuclear Data Tables, 40, 283  

\bibitem[Chevalier \& Soderberg(2010)]{2010ApJ...711L..40C} Chevalier, R.~A., \&
  Soderberg, A.~M.\ 2010, \apjl, 711, L40   

\bibitem[Claeys et al.(2011)]{2011A&A_528A_131C} Claeys, J.~S.~W., de
  Mink, S.~E., Pols, O.~R., Eldridge, J.~J., \& Baes, M.\ 2011, \aap,
  528, A131   

\bibitem[Clocchiatti et al.(1996)]{1996ApJ...462..462C} Clocchiatti, A., 
Wheeler, J.~C., Brotherton, M.~S., et al.\ 1996, \apj, 462, 462 

\bibitem[de Jager et al.(1988)]{1988A&AS_72_259D} de Jager, C.,
  Nieuwenhuijzen, H., \& van der Hucht, K.~A.\ 1988, \aaps, 72, 259   

\bibitem[De Vito \& Benvenuto(2012)]{2012MNRAS.tmp.2583D} De Vito,
  M.~A., \& Benvenuto, O.~G.\ 2012, \mnras, 2583   

\bibitem[Dessart et al.(2011)]{2011MNRAS.414.2985D} Dessart, L., Hillier, 
D.~J., Livne, E., et al.\ 2011, \mnras, 414, 2985  

\bibitem[Eldridge et al.(2008)]{2008MNRAS.384.1109E} Eldridge, J.~J., 
Izzard, R.~G., \& Tout, C.~A.\ 2008, \mnras, 384, 1109 

\bibitem[Elias-Rosa et al.(2009)]{2009ApJ...706.1174E} Elias-Rosa, N., Van 
Dyk, S.~D., Li, W., et al.\ 2009, \apj, 706, 1174  

\bibitem[Elias-Rosa et al.(2010)]{2010ApJ...714L.254E} Elias-Rosa, N., Van 
Dyk, S.~D., Li, W., et al.\ 2010, \apjl, 714, L254  


\bibitem[Eggleton(1983)]{1983ApJ_268_368E} Eggleton, P.~P.\ 1983, \apj,
268, 368  

\bibitem[Filippenko(1997)]{1997ARA&A..35..309F} Filippenko,
  A.~V.\ 1997, \araa, 35, 309 

\bibitem[Fraser et al.(2010)]{2010ApJ...714L.280F} Fraser, M., Tak{\'a}ts, 
K., Pastorello, A., et al.\ 2010, \apjl, 714, L280  


\bibitem[Georgy et al.(2009)]{2009A&A...502..611G} Georgy, C., Meynet,
  G., Walder, R., Folini, D., \& Maeder, A.\ 2009, \aap, 502, 611  

\bibitem[Georgy(2012)]{2012A&A_538L_8G} Georgy, C.\ 2012, \aap, 538, L8  

\bibitem[Heger et al.(2003)]{2003ApJ...591..288H} Heger, A., Fryer,
  C.~L.,  Woosley, S.~E., Langer, N., \& Hartmann, D.~H.\ 2003, \apj,
  591, 288  

\bibitem[Kippenhahn et  al.(1980)]{1980A&A_91_175K} Kippenhahn, R.,
Ruschenplatt, G., \& Thomas, H.-C.\ 1980, \aap, 91, 175  

\bibitem[Kurucz(1993)]{1993sssp.book.....K} Kurucz, R.~L.\ 1993, Kurucz 
CD-ROM, Cambridge, MA: Smithsonian Astrophysical Observatory, |c1993, 
December 4, 1993  
  
\bibitem[Langer et  al.(1983)]{1983A&A_126_207L} Langer, N., Fricke, K.~J., \&
Sugimoto, D.\ 1983, \aap, 126, 207  

\bibitem[Langer et al.(1985)]{1985A&A_145_179L} Langer, N., El Eid, M.~F., \&
Fricke, K.~J.\ 1985, \aap, 145, 179  

\bibitem[Maund et al.(2004)]{2004Natur_427_129M} Maund, J.~R., Smartt, 
S.~J., Kudritzki, R.~P., Podsiadlowski, P., 
\& Gilmore, G.~F.\ 2004, \nat, 427, 129  

\bibitem[Maund et al.(2011)]{2011ApJ_739L_37M} Maund, J.~R., Fraser,
  M., Ergon, M., et al.\ 2011, \apjl, 739, L37  

\bibitem[Mauron \& Josselin(2011)]{2011A&A...526A.156M} Mauron, N., \&
  Josselin, E.\ 2011, \aap, 526, A156  

\bibitem[Neo et al.(1977)]{1977PASJ_29_249N} Neo, S., Miyaji, S., 
Nomoto, K., \& Sugimoto, D.\ 1977, \pasj, 29, 249  

\bibitem[Nomoto et al.(1993)]{1993Natur_364_507N} Nomoto, K., Suzuki, T., 
Shigeyama, T., et al.\ 1993, \nat, 364, 507  

\bibitem[Podsiadlowski et al.(1993)]{1993Natur_364_509P} Podsiadlowski, 
P., Hsu, J.~J.~L., Joss, P.~C., \& Ross, R.~R.\ 1993, \nat, 364, 509 

\bibitem[Podsiadlowski et al.(2002)]{2002ApJ_565_1107P} Podsiadlowski,  P.,
Rappaport, S., \& Pfahl, E.~D.\ 2002, \apj, 565, 1107  

\bibitem[Ritter(1988)]{1988A&A_202_93R} Ritter, H.\ 1988, \aap, 202,  93  

\bibitem[Ryder et al.(2006)]{2006MNRAS.369L..32R} Ryder, S.~D., Murrowood, 
C.~E., \& Stathakis, R.~A.\ 2006, \mnras, 369, L32  

\bibitem[Shigeyama et al.(1994)]{1994ApJ...420..341S} Shigeyama, T., 
Suzuki, T., Kumagai, S., et al.\ 1994, \apj, 420, 341 

\bibitem[Smith et al.(2011)]{2011MNRAS.412.1522S} Smith, N., Li, W., 
Filippenko, A.~V., \& Chornock, R.\ 2011, \mnras, 412, 1522  

\bibitem[Soderberg et al.(2012)]{2012ApJ...752...78S} Soderberg, A.~M., 
Margutti, R., Zauderer, B.~A., et al.\ 2012, \apj, 752, 78  

\bibitem[Stancliffe 
\& Eldridge(2009)]{2009MNRAS.396.1699S} Stancliffe, R.~J., \&
  Eldridge, J.~J.\ 2009, \mnras, 396, 1699  

\bibitem[Tak{\'a}ts \& Vink{\'o}(2006)]{2006MNRAS.372.1735T}
  Tak{\'a}ts, K., \& Vink{\'o}, J.\ 2006, \mnras, 372, 1735  

\bibitem[van Loon et al.(2005)]{2005A&A...438..273V} van Loon, J.~T.,
  Cioni, M.-R.~L., Zijlstra, A.~A., \& Loup, C.\ 2005, \aap, 438, 273  

\bibitem[Van Dyk et al.(2011)]{2011ApJ_741L_28V} Van Dyk, S.~D., Li,
  W., Cenko, S.~B., et al.\ 2011, \apjl, 741, L28  

\bibitem[Woosley et al.(1994)]{1994ApJ...429..300W} Woosley, S.~E., 
Eastman, R.~G., Weaver, T.~A., \& Pinto, P.~A.\ 1994, \apj, 429, 300 

\bibitem[Yoon et al.(2010)]{2010ApJ_725_940Y} Yoon, S.-C., Woosley, 
S.~E., \& Langer, N.\ 2010, \apj, 725, 940  

\end{thebibliography}
\end{document}